\documentstyle[twocolumn,aps,prl]{revtex}

\input epsf.sty
\input rotate.tex

\begin{document}

\wideabs{
\draft
\title{Description of the RHIC $p_\perp$-spectra in a thermal model with
expansion \cite{grant}}
\author{Wojciech Broniowski and Wojciech Florkowski}
\address{The H. Niewodnicza\'nski Institute of Nuclear Physics,
PL-31342 Cracow, Poland}
\maketitle
\begin{abstract}
The assumption of simultaneous chemical and thermal freeze-outs 
of the hadron gas leads to a surprisingly accurate, albeit 
entirely conventional, explanation of the
recently-measured RHIC $p_\perp$-spectra.
The original thermal spectra are supplied with secondaries from 
cascade decays
of all resonances, and subsequently folded with a suitably 
parameterized expansion involving longitudinal and transverse flow. 
The predictions of this thermal approach, with various
parametrizations for the expansion, are in a striking
quantitative agreement with the data in the whole available range of $0 \le 
p_\perp \le 3.5$GeV.
\end{abstract}
}

In this Letter we offer a very simple explanation of the $p_{\perp
}$-spectra recently measured at RHIC \cite{velko,harris,QM}. Our
approach has the following ingredients: i) simultaneous  {\em chemical} and 
{\em thermal} freeze-outs, with the hadron distributions given by the thermal
model; in other words, hadrons decouple completely when
the thermodynamic parameters reach the freezing conditions, and no 
particle rescattering after freeze-out is present, ii) these 
thermal distributions are folded with a suitably parameterized
hydrodynamic expansion, involving longitudinal and transverse flow,
finally, iii) feeding from resonances, including cascades, 
is incorporated in a complete way.

So far, the thermal approach has been applied successfully in studies of
particle ratios measured in relativistic heavy-ion collisions at AGS
and SPS \cite{pbmags,raf,cest,pbmsps,yg,becatt,gaz}. Quite recently, it has
also been shown that particle ratios measured at RHIC may be
equally well described in the framework of such models
\cite{raf0,pbmrhic,fbm}. Description of hadronic $p_{\perp}$-spectra in
thermal models is more involved, since the spectra are affected by
decays of resonances, hydrodynamic flow, and possibly by other
phenomena occurring during the alleged phase transition from the
quark-gluon plasma to a hadron gas \cite {dfh}.
The model results presented in this Letter are in a surprising
agreement with the experiment in the entire range of the data, $0 \le
p_\perp \le 3.5$GeV, as can be seen in Fig. 1.  The model has two free
parameters: one controlling the size of the system
(overall normalization of the spectra), and the other one the
transverse flow.  We test two different models (parametrizations) for
the freeze-out hypersurface and the hydrodynamic expansion.  Both
combine the Bjorken expansion \cite{bjorken} with transverse flow
\cite{baym,Kolya}, and follow the spirit of
Refs. \cite{siemens,SSH,BL,Rischke,SH}.

The first model (model I) assumes that the freeze-out
takes place at a fixed value of
the invariant time,
$ \tau =\sqrt{t^{2}-r_{z}^{2}-r_{x}^{2}-r_{y}^{2}}=\hbox{const}$,
which means that, due to time dilation,
the particles in the fluid elements moving farther away
from the collision center decouple later than the particles in the fluid
elements remaining at rest in the center-of-mass system of the colliding
nuclei. Furthermore, we assume that
the four-velocity of expansion is proportional to the coordinate,
\begin{equation}
u^{\mu } =\frac{x^{\mu }}{\tau }=\frac{t}{\tau }\left(
1,\frac{r_{z}}{t},\frac{r_{x}}{t},\frac{r_{y}}{t}\right).  \label{umu}
\end{equation}
The freeze-out hypersurface is parameterized as \cite{BL}
\begin{eqnarray}
t &=&\tau\, {\rm ch}  \alpha _{\parallel }\,{\rm ch}  \alpha _{\perp },
\quad r_{z}=\tau\,
{\rm sh}  \alpha _{\parallel }\,{\rm ch}  \alpha _{\perp },  \nonumber \\
r_{x} &=&\tau\, {\rm sh}  \alpha _{\perp }\cos \phi ,
\quad r_{y}=\tau\, {\rm sh}  \alpha
_{\perp }\sin \phi ,  \label{par}
\end{eqnarray}
where $\alpha _{\parallel }$ is the rapidity of the fluid element ($%
v_{z}=r_z/t=\tanh \alpha _{\parallel }$), whereas $\alpha _{\perp }$ describes
the transverse size of the system 
($\rho =\tau\, {\rm sh}  \alpha _{\perp }$).
The transverse velocity is
$v_\rho=\tanh \alpha_\perp/{\rm ch}  \alpha_\parallel$.
We account for the finite transverse size of the system by imposing
the condition $\rho =\sqrt{r_{x}^{2}+r_{y}^{2}}<\rho _{max}$.
The second model considered (model II) has the same 
parametrization of the four-velocity and the freeze-out
hypersurface as the popular {\em blast}
model \cite{SSH,Rischke}:
\begin{eqnarray}
 u^{\mu } &=& \left(
{\rm ch}  \alpha_{\parallel} {\rm ch}  \beta_\perp,
{\rm sh}  \alpha_{\parallel} {\rm ch}  \beta_\perp, \cos \phi {\rm sh}  \beta_\perp,
\sin \phi {\rm sh}  \beta_\perp \right),  \nonumber \\
 t &=& \tau {\rm ch}  \alpha _{\parallel },\quad r_{z}=\tau
{\rm sh}  \alpha _{\parallel },  \nonumber \\
 r_{x} &=& \tau {\rm sh}  \alpha _{\perp }\cos \phi ,\quad r_{y}=\tau {\rm sh}  \alpha
_{\perp }\sin \phi .  \label{parbl}
\end{eqnarray}
Both models I and II are boost-invariant. Since deviations from
boost-invariance are seen in the rapidity distributions at RHIC
\cite{QM}, our present approach should be regarded as an approximate
treatment of the mid-rapidity region. However, this approximation is
very good. One may depart from the boost-invariance by limiting the
integration in the $\alpha_\parallel$ variable, thus limiting the
longitudinal size of the system.  The results for the spectra obtained
that way are very similar to the boost-invariant results presented
below even for the limit for $\alpha_\parallel$ as low as 0.5.

The local freeze-out conditions, {\em i.e.}, the values of the
temperature and the chemical potentials, are universal for the whole
freeze-out hypersurface. Since for boost-invariant
models the particle ratios at mid-rapidity are not
affected by the expansion (this important
point is discussed below), the values of the thermodynamic parameters may be
obtained directly 
from the standard thermal analysis, which yields \cite{fbm} 
$T=165\pm 7{\rm MeV}$, $\mu_{B}=41\pm 5{\rm MeV}$, $\mu_{S}=9{\rm MeV}$, and 
$\mu_{I}=-1{\rm MeV}$. 
Knowing $T$
and $\mu$'s we calculate the local distribution functions of hadrons which
include the {\em initial thermal contribution}, as well as additional
contributions from the {\em sequential two- and three-body 
decays of {\em all} heavier resonances}.
These decays are very important, since they {\em effectively cool} 
the system by
35-40 MeV, as recently shown in Ref. \cite{fbm}, and also known from earlier
works on other reactions \cite{skh,bsw}. 
The standard Cooper-Frye-Schonberg formula \cite{cfs} is used to
calculate the $p_{\perp }$-spectra of the observed hadrons.
As the result, the particle densities are obtained as the integrals 
over the freeze-out
hypersurface, 
$N_{i}=\int d^{3}p /p^{0}\int p^{\mu }d\Sigma_{\mu} f_{i}(p\cdot u)$, 
where $d\Sigma _{\mu }$ is the volume element of
the hypersurface, $f_{i}$ is the phase-space distribution
function for particle species $i$ (composed from the initial and secondary
particles), and 
$p^{\mu }=\left( m_{\perp }{\rm ch}y, m_{\perp }{\rm sh}y, 
p_{\perp }\cos \varphi ,p_{\perp }\sin \varphi \right)$ is the four-momentum. 
Finally, we find, for the 
case of model I, the
rapidity and transverse-momentum distributions of hadrons, 
\begin{eqnarray}
\frac{dN_{i}}{d^{2}p_{\perp }dy} &=&\ \tau ^{3}\int_{-\infty }^{+\infty
}d\alpha _{\parallel }\int_{0}^{\rho _{\max }/\tau }{\rm sh}  \alpha _{\perp
}d\left( {\rm sh}  \alpha _{\perp }\right)  \nonumber \\
&&\times \int_{0}^{2\pi }d\xi \ p\cdot u\ f_{i}\left( p\cdot u\right) ,
\label{dNi}
\end{eqnarray}
where 
$p\cdot u=m_{\perp }{\rm ch}  \left( y-\alpha _{\parallel }\right) {\rm ch}  \alpha
_{\perp }-p_{\perp }\cos \xi \, {\rm sh}  \alpha _{\perp } $
and $\xi =\phi -\varphi $.
Similarly, for model II the relevant parameters are the velocity, 
$\beta_\perp$, and the volume of the system \cite{SSH,Rischke}.
We observe that the rapidity distribution (\ref
{dNi}) 
is boost-invariant, since the dependence on $y$ can be absorbed in
the integration variable by shifting $\alpha _{\parallel }\rightarrow \alpha
_{\parallel }+y.$ Clearly, this is a direct consequence of the assumed
boost-invariant form of the freeze-out surface.  

In the thermal-model fit of particle ratios of Ref. \cite{fbm} 
one computes the integrals $N_{i}=V \int d^{3}p\
f_{i} ( \sqrt{m_{i}^{2}+p^{2}}) $. The question arises whether the
ratios obtained that way, which correspond to the collection 
of particles from the whole phase space, 
are the same as the ratios
experimentally measured in the mid-rapidity region, {\em i.e.,} the ratios
of the integrals $dN_{i}/dy=\int d^{2}p_{\perp }dN_{i}/(d^{2}p_{\perp }dy)$.
The answer is yes, and follows from the boost-invariance of the expansion
model. Indeed, since $dN_{i}/dy$ are independent of $y$, we have 
\begin{equation}
\frac{dN_{i}/dy}{dN_{j}/dy}=\frac{\int dy\,dN_{i}/dy}{\int dy\,dN_{j}/dy}=%
\frac{N_{i}}{N_{j}}.  \label{general}
\end{equation}
This obvious general result can be verified explicitly in our specific
boost-invariant models. 

Figure 1 shows our main result. In Fig. 1 (a,b) we compare the model 
predictions for the $p_{\bot }$-spectra of pions, kaons, 
protons and antiprotons, with the PHENIX 
{\em minimum bias} preliminary data \cite{velko}.
The model parameters are fitted by the 
$\chi^2$ method including all points in the range $0 < p_\perp < 2$GeV 
from Fig. 1 (a,b). This yields $\tau=5.55$fm, $\rho_{\rm max}=4.50$fm for model I, 
and ${\cal V}=(6.48{\rm fm})^3$, $\beta_\perp=0.52$ for model II. 
The statistical errors in the fitted parameters are of the order of 1\%.  
In Fig. 1 (c) we compare the model predictions to preliminary PHENIX
\cite{velko} and STAR \cite{harris}
data for the {\em most central} collisions. 
The model parameters are fitted by the least-square method to all 
points in Fig. 1 (c), yielding $\tau=7.66$fm, $\rho_{\rm max}=6.69$fm 
for model I, and ${\cal V}=(9.81 {\rm fm})^3$, $\beta_\perp=0.52$ for model II.
It is natural that the size parameters of the system 
for the most central collisions are larger
than for the minimum-bias case, which averages over centralities. 
The size parameters correspond to the size of the system, in particular 
for central collisions in model I $\rho_{\rm max}=6.69$fm 
is very close to the radius of the Au 
nucleus, 6.22fm. 
We note that the quality of the fit in Fig. 1 is impressive. For the 
minimum-bias data (Fig. 1 (a,b))
model I
(thicker lines) crosses virtually all data points 
within error bars, with $\chi^2$/degree of freedom less than 1. Amusingly, 
also the very-high $p_\perp$ data are reproduced. The fit with model II 
is very similar in the range $0 < p_\perp < 2$GeV, and falls 
slightly below
the data at higher $p_\perp$, where
hard processes are expected to contribute.
The fit to the most central collisions, Fig 1 (c), is of similar quality
except for the $\overline{p}$ preliminary 
data from STAR, which fall 30-50\% below the 
model fits at low $p_\perp$. Note, however, that at low $p_\perp$ 
the STAR and PHENIX preliminary data are not fully consistent. 
The total multiplicity of charged particles produced in the 
most central event measured at PHOBOS \cite{phobos}, $555 \pm 12 \pm 35$, 
is consistent with size parameters quoted above, and yields $\tau \simeq 7$fm
for model I and ${\cal V} \simeq (9 {\rm fm})^3$ for model II.

Since the values of the strange and isospin
chemical potentials are very close to zero, the model predictions for $\pi
^{+}$ and $\pi ^{-}$, as well as for $K^{+}$ and $K^{-}$ are practically the
same. The value of the baryon chemical potential of $41$ MeV splits the $p$
and $\bar{p}$ spectra. Note the convex shape of the pion spectra in Fig. 1,
reproduced by the model. In addition, 
the $\pi^+$ and $p$ curves in Fig. 1 cross at $p_\perp \simeq 2$ GeV, 
and the $K^+$ and $p$ at $p_\perp \simeq 1$ GeV, exactly as in the
experiment. We stress that our method is different from traditional
fits in the blast or similar models, where the temperatures for 
various particles are being adjusted independently or from
semi-empirical formulas. We have no freedom here:
the temperature is the freeze-out temperature fixed by the particle ratios, 
and the spectra are obtained as described in Ref. \cite{fbm}.

In Table 1 we present the inverse slope parameters, $T_{\rm eff}$, defined by
fitting the function $const \exp(-m_\perp/T_{\rm eff})$ to the data
\footnote{\label{foot} Note that this definition depends on 
the fitting region in $m_\perp$. Thus, the resulting numbers depend 
quite strongly on this region, and $T_{\rm eff}$ is a largely-biased 
measure of the particle spectra. 
Following Ref. \cite{velko}, we use 
$0.3 {\rm GeV} < p_\perp < 0.9 {\rm GeV}$
for the pions,  $0.55 {\rm GeV} < p_\perp < 1.6 {\rm GeV}$ for $K^+$, $p$, and
$\overline{p}$, and  $0.75 {\rm GeV} < p_\perp < 1.6 {\rm GeV}$ for $K^-$. The 
$\chi^2$ fits are performed in these regions.}, and average $p_\perp$. 
Agreement of model I and 
the data for the inverse slopes 
is within error bars except for most central $\overline{p}$ data 
from PHENIX (2.5 standard deviations) and $K^-$ and $\overline{p}$ data
from STAR 
(2 and 4.6 standard deviations, respectively).
% This may indicate problems in the preliminary STAR data for $\overline{p}$. 
The values of $\langle p_\perp \rangle$ are within error bars. 

We end this Letter with a more pedagogical discussion of the role of various
effects included in our analysis. In Fig. 2 the dotted line shows the 
initial pion $p_{\perp }$-spectrum in a static fireball with the same temperature and
chemical potentials as used in our calculation. No secondaries are
included here. The effect of the decays of {\it all}
resonances is represented by the dashed-dotted line, representing the sum of
the initial and secondary pions. Decays of the
resonances lead to an {\em effective decrease of the temperature} 
by about 35-40
MeV \cite{fbm}, since the emitted particles tend to populate the low-$%
p_{\perp \text{ }}$region. However, the spectrum remains concave. The effect
of the pure longitudinal Bjorken expansion (with $\tau =\sqrt{t^{2}-r_{z}^{2}%
}={\rm const}$) is illustrated by the dashed line. This is a {\em redshift} effect,
since all fluid elements move away from the observer, which leads to extra
cooling of the spectrum. The solid line corresponds to model I (model II 
gives very similar results),
incorporating both the longitudinal expansion and the transverse flow. The
transverse flow causes some fluid element to move in the direction of the
observer, leading to {\em blueshift }\cite{heinz}. Hence we find a
combination of redshift and blueshift, yielding the $p_{\bot }$ spectrum
displayed by the solid line in Fig. 3. Note that the spectrum finally
acquires the convex shape, as seen in the experiment ({\em cf.} Fig. 1).
The effects of blueshift are stronger for more massive
particles, hence the behavior of Fig. 1 and Table 1.

To conclude, we emphasize that the presented 
description, implementing in a
simple fashion all key ingredients: freeze-out, decays of resonances, and
longitudinal and transverse flows, works for RHIC in the whole 
available range of data. We stress that we 
have assumed that the
chemical and thermal freeze-outs occur simultaneously, 
which is in the spirit of the sudden hadronization of Refs.
\cite{raf1,raf2}. 
A practical value of our results is that they give 
hints and constraints for more 
involved hydrodynamic calculations, {\em e.g. }
\cite{shur,kolb,huovinen},
by providing the freeze-out conditions that describe the data. 
A natural extensions of the model should include different 
centrality effects (elliptic flow), and rapidity dependence.
Also, the model must be further verified against the available data from the 
HBT pion interferometry \cite{star}. Note that our size parameters are 
very similar to the experimental HBT radii.   
We have checked that our model works also for the SPS data. 
The details of these studies will be presented elsewhere.

We are grateful to Ulrich Heinz for extensive and useful discussions.

\newpage
~
\newpage
~

\widetext

\newsavebox{\lala}

\savebox{\lala}{\vbox{
\begin{figure}[h]
\epsfysize=14cm
\centerline{\mbox{\epsfbox{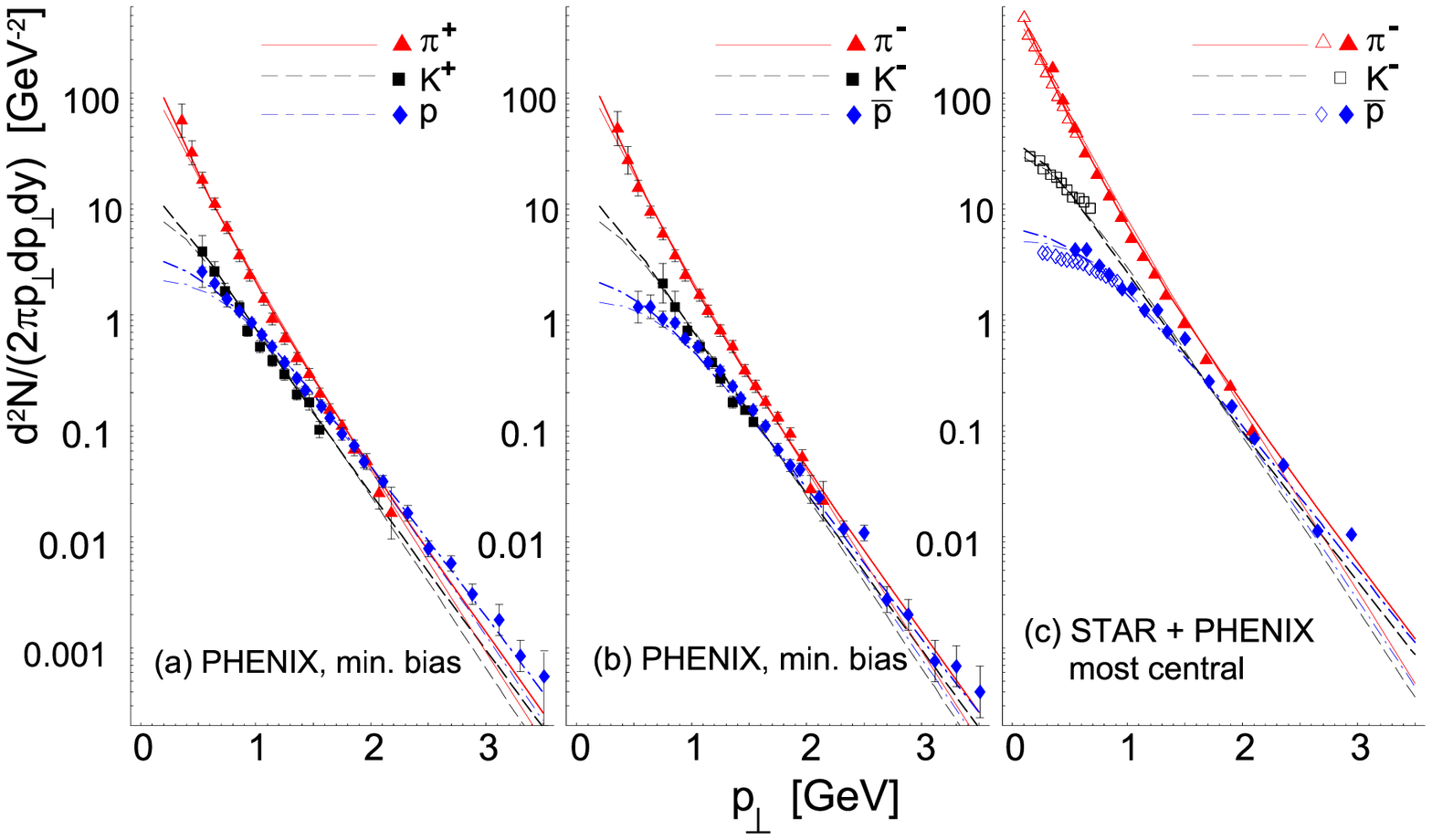}}}
\caption{The $p_{\perp}$-spectra at $y=0$
of pions (solid line), kaons (dashed line)
and protons or antiprotons (dashed-dotted line), as evaluated from model I
(thicker lines) and model II (thinner lines), compared to
the PHENIX preliminary minimum-bias data (a,b), and
to STAR (open symbols) and PHENIX preliminary highest-centrality data (c)
($Au+Au$ at %
% $\sqrt{s}=
130GeV
).}
\label{f1}
\end{figure}
}}

~
\vspace{1.7cm}
~

\centerline{\rotl{\lala}}

\newpage
~
\newpage
\narrowtext

\begin{figure}[h]
\epsfysize=6.5cm
\centerline{\mbox{\epsfbox{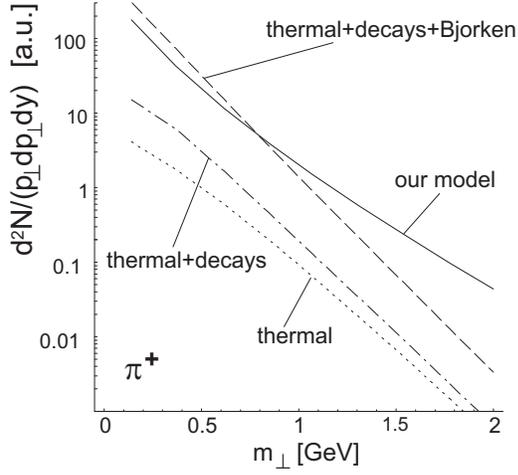}}}
\caption{Contributions of various effects to the $p_\perp$-spectra of
$\pi^+$ (normalizations arbitrary). }
\label{f3}
\end{figure}

\begin{table}[b]
\caption{Inverse slope parameters, $T_{\rm eff}$,
and the average transverse momentum, $ \langle p_\perp \rangle$.}
\begin{tabular}{r|d|d|d}
 $T_{\rm eff}$ [MeV]  & $\pi$ & $K$ &
$p$ or $\overline{p}$ \\
\tableline
PHENIX, min. bias, \hfill ~~exp. & $190 \pm 17$ & $272 \pm 16 $ & $274 \pm 11$  \\
positive hadrons \hfill ~~model I  & 203 & 263 & 292 \\
\tableline
PHENIX, min. bias, \hfill ~~exp. & $207 \pm 20$ & $260 \pm 17 $ & $301 \pm 14$  \\
negative hadrons \hfill ~~model I  & 202 & 255 & 291 \\
\tableline
PHENIX, most central, \hfill ~~exp. & $197 \pm 19$ & -  & $367 \pm 23$  \\
negative hadrons \hfill ~~model I  & 206 & 273  & 310 \\
\tableline
STAR, most central, \hfill ~~exp. & $188 \pm 20$ & $300 \pm 35 $ & $560 \pm 50$  \\
negative hadrons \hfill ~~model I      & 176 & 223 & 327 \\
\tableline
\tableline
$ \langle p_\perp \rangle$ [MeV]
&  $\pi$  & $K$ &  $p$ or $\overline{p}$  \\
\tableline
PHENIX, most central, \hfill ~~exp. & $370 \pm 70$ & $600 \pm 60$ & $840 \pm 50$   \\
positive hadrons  \hfill ~~ model I  & 438 & 629 & 853  \\
\tableline
PHENIX, most central, \hfill ~~exp. & $370 \pm 70$ & $630 \pm 80$ & $860 \pm 50$   \\
negative hadrons \hfill ~~ model I  & 434 & 629 & 852  \\
\end{tabular}
\end{table}


\begin{references}

\bibitem[\ast]{grant} Supported in part by the Polish State Committee for 
Scientific Research, grant 2 P03B 09419.

\bibitem{velko}  J. Velkovska for the PHENIX Collaboration, 
nucl-ex/0105012; A. Bazilevsky for the PHENIX Collaboration, nucl-ex/0105017.

\bibitem{harris} J. W. Harris for the STAR Collaboration, in
\cite{QM}.

\bibitem{QM} Proceedings of the {\em Quark Matter 2001} conference,
BNL, January 2001, Nucl. Phys. A (in print).

\bibitem{pbmags}  P. Braun-Munzinger, J. Stachel, J. P. Wessels, and N. Xu,
Phys. Lett. B {\bf 344}, 43 (1995); Phys. Lett. B {\bf 365}, 1 (1996).

\bibitem{raf} J. Rafelski, J, Letessier, and A. Tounsi,
Acta Phys. Pol. B {\bf 28}, 2841 (1997).

\bibitem{cest}  J. Cleymans, D. Elliott, H. Satz, and R. L. Thews, Z. Phys.
C {\bf 74}, 319 (1997).

\bibitem{pbmsps}  P. Braun-Munzinger, I. Heppe, and J. Stachel, Phys. Lett.
B {\bf 465}, 15 (1999).

\bibitem{yg}  G. D. Yen and M. I. Gorenstein, Phys. Rev. C {\bf 59}, 2788
(1999).


\bibitem{becatt}
F. Becattini, J. Cleymans, A. Keranen, E. Suhonen, and K. Redlich,
Phys. Rev. C {\bf 64}, 024901 (2001). 


\bibitem{gaz} M. Ga\'zdzicki, Nucl. Phys. A {\bf 681}, 153 (2001).

\bibitem{raf0} J. Rafelski, J. Letessier, and G. Torrieri,
Phys. Rev. C {\bf 64}, 054907 (2001).

\bibitem{pbmrhic}  P. Braun-Munzinger, D. Magestro, K. Redlich, and J.
Stachel, hep-ph/0105229.

\bibitem{fbm}  W. Florkowski, W. Broniowski, and M. Michalec,
nucl-th/0106009.

\bibitem{dfh}  J. Dolej\v s\'{\i}, W. Florkowski, and J. H\"ufner, Phys.
Lett. B {\bf 349}, 18 (1995).

\bibitem{bjorken} J. D. Bjorken, Phys. Rev. D {\bf 27}, 140 (1983).

\bibitem{baym} G. Baym, B. Friman, J.-P. Blaizot, M. Soyeur, and W. Czy\.z,
Nucl. Phys. A {\bf 407}, 541 (1983). 

\bibitem{Kolya} P. Milyutin and N. N. Nikolaev, Heavy Ion Phys {\bf 8}, 
333 (1998); V. Fortov, P. Milyutin, and N. N. Nikolaev,
JETP Lett. {\bf 68}, 191 (1998).

\bibitem{siemens} P. J. Siemens and J. Rasmussen, Phys. Rev. Lett. {\bf 42},
880 (1979); P. J. Siemens and J. I. Kapusta, Phys. Rev. Lett. {\bf 43},
1486 (1979).

\bibitem{SSH} E. Schnedermann, J. Sollfrank, and U. Heinz, Phys. Rev. C {\bf 48},
2462 (1993).

\bibitem{BL} T. Cs\"{o}rg\H{o} and B. L\"{o}rstad, Phys. Rev. C {\bf 54}, 1390
(1996).

\bibitem{Rischke} D. H. Rischke and M. Gyulassy, Nucl. Phys. A {\bf 697}, 701 
(1996); Nucl. Phys. A {\bf 608}, 479 (1996).


\bibitem{SH} R. Scheibl and U. Heinz, Phys. Rev. C {\bf 59}, 1585 (1999).

\bibitem{skh} J. Sollfrank, P. Koch, and U. Heinz, Phys. Lett. B {\bf 252},
256 (1990).

\bibitem{bsw} G. E. Brown, J. Stachel, and G. M. Welke, Phys. Lett. B
{\bf 253}, 19 (1991).

\bibitem{cfs}  F. Cooper, G. Frye, and E. Schonberg, Phys. Rev. D {\bf 11},
192 (1975).

\bibitem{jpg}  J. Cleymans, H. Oeschler, and K. Redlich, J. Phys. G {\bf 25}%
, 281 (1999).

\bibitem{phobos} B. B. Back {\em et al.}, PHOBOS Collaboration,
Phys. Rev. Lett. {\bf 85}, 3100 (2000). 

\bibitem{heinz}  U. Heinz, Nucl. Phys. A {\bf 661}, 140 (1999).

\bibitem{raf1} J. Rafelski and J. Letessier, Phys. Rev. Lett. {\bf 85}, 4695
(2000).

\bibitem{raf2} G. Torrieri and J. Rafelski, New Jour. Phys. {\bf 3}, 12 (2001).


\bibitem{shur} D. Teaney, J. Lauret, and E. V. Shuryak,
Phys. Rev. Lett. {\bf 86}, 4783 (2001);  
nucl-th/0104041, in \cite{QM}.


\bibitem{kolb} P. Huovinen, P. F. Kolb, U. Heinz, P. V. Ruuskanen, and 
S. A. Voloshin, Phys. Lett. B {\bf 503}, 58 (2001).

\bibitem{huovinen} P. Huovinen, nucl-th/0108033.


\bibitem{star} C. Adler et al., STAR Collaboration, Phys. Rev. Lett. {\bf 87},
082301 (2001).


\end{references}
\end{document}